\documentclass[12pt]{iopart}
\usepackage{iopams} 
\usepackage{graphicx} 
\usepackage{amssymb}
\begin{document}

\title{Energy exchange and transition to localization in the asymmetric Fermi-Pasta-Ulam oscilliatory chain}

\author {Smirnov V.V., Shepelev D.S., Manevitch L.I.}

\ead{deshepelev@gmail.com}

\address{N.N.Semenov Institute of Chemical Physics, RAS, 4 Kosygin str., 119991, Moscow, Russia}

\begin{abstract}
A finite (periodic) FPU chain is chosen as a convenient point for investigating the energy exchange phenomenon in nonlinear oscillatory systems. 
As we have recently shown, this phenomenon may occur as a consequence of the resonant interaction between high-frequency nonlinear normal modes. 
This interaction determines both the complete energy exchange between different parts of the chain and the transition to energy localization in an excited group of particles. 
In the paper, we demonstrate that this mechanism can exist in realistic (asymmetric) models of atomic or molecular oscillatory chains. 
Also, we study the resonant interaction of conjugated nonlinear normal modes and prove a possibility of linearization of the equations of motion. 

The theoretical constructions developed in this paper are based on the concepts of "effective particles" and Limiting Phase Trajectories. 
In particular, an analytical description of energy exchange between the "effective particles" in the terms of non-smooth functions is presented. 
The analytical results are confirmed with numerical simulations.
\end{abstract}



\maketitle


\section{Introduction}

It is well known, that the problem of vibration energy localization in the infinite Fermi-Pasta-Ulam (FPU) chain can be asymptotically reduced to finding a localized solution (breather) of the continuum Nonlinear Schro\-din\-ger Equation (NLSE), which is a completely integrable system \cite{Scott}.
After such reduction to NLSE, the mathematical origin of the phenomenon becomes clear.
However, it was recently shown that not an infinite but a finite (periodic) $\beta$-FPU chain has to be chosen to clarify the physical nature of vibration energy localization \cite{PhysRevE}.
This model has been intensively studied over the last decade (\cite{PoggiRuffo}, \cite{Rink}, \cite{Henrici}, \cite{Flach}, \cite{Dauxois}) in the framework of Nonlinear Normal Modes (NNMs) concept \cite{ManMikhl}, \cite{Vakakis}.
On the contrary to NLSE, finite discrete systems are not completely integrable (except the three particle $\beta$-FPU periodic model, \cite{Feng}).
Therefore, the knowledge of NNMs leads to a unique possibility for constructing a wide class of solutions using a perturbation techniques \cite{Scott}.
We have recently shown that an increase of the particles number in the symmetric ( $\beta$-FPU) chain leads inevitably to the resonant interaction between the high frequency NNMs because of the linear spectrum densification. 
We suggested the concepts of "effective particles"\ and Limiting Phase Trajectories (LPTs) to understand and describe both complete energy exchange between different parts of the chain and transition to energy localization \cite{PhysRevE}. 
From the modal viewpoint, such exchange results from the resonant interaction of the zone-boundary mode ($\pi$-mode) with its nearest neighbours.
In this case, the formation of a mobile localized mode is accompanied by a prohibition of the transfer of energy from one part of the system to another. 
The complete energy exchange between "effective particles", which is described using LPT concept \cite{ArchApplMech}, is an alternative to nonlinear normal oscillations characterized by the conservation of energy. 

The symmetric model of the FPU chain is valid for various mechanical problems. 
However, for applications to the problems of Molecular and Polymer Physics it is necessary to consider the asymmetric ($\alpha-\beta$) FPU problem because realistic interatomic potentials of interaction are strongly asymmetric.
The presence of the asymmetric term in the interaction potential leads to significant complication of the problem. 
First of all, the NNMs of the discrete FPU chain with periodic boundary conditions cannot be divided into closed 4-manifolds, as it is possible in the case of the $\beta$-FPU chain \cite{PhysRevE}. 
The modal equations for the degenerate pairs of modes corresponding to the same eigenvalue are valid only for special initial conditions. 

One of the main points under consideration is whether the presence of the quad\-ra\-tic terms in the equations of motion leads to an ef\-fec\-ti\-ve renormalization of the constant of the fourth-power interaction. 
Obtaining the positive answer, we present an analytical description of  energy exchange between "effective particles" in the ($\alpha - \beta $)-FPU chain using the LPT concept.

Note that the same effective renormalization of the quartic coupling constant was introduced for modulational instability analyses of modes close to the zone-boundary mode in the rotating-wave approximation (\cite{Sandusky}, \cite{Burlakov}). The current paper considers a more wide class of problems, a the key result of this paper is in an analytical description of intensive energy exchange between "the effective particles" and energy localization on the effective particle in terms of the non-smooth basic functions.

We also study the asymmetry effect on the threshold to energy localization.
In conclusion, the comparison of analytical solutions with the results of numerical simulation is performed.

\section{The model and the equations of motion.}

We consider the $(\alpha-\beta)$-FPU system defined by the Hamilton function
$$H_0=\sum\limits_{j=1}^{N}\frac{1}{2}p_{j}^{2}+\frac{1}{2}( q_{j+1}-q_{j} )^2+\frac{\alpha}{3}( q_{j+1}-q_{j} )^3 + \frac{\beta}{4}( q_{j+1}-q_{j} )^4\eqno(1)$$
and the periodic boundary conditions $(q_{N+1}=q_{1}, p_{N+1}=p_{1})$, where $q_j$ and $p_j$ are the coordinates and the conjugate momenta, respectively, $N$ is the number of particles.

The transformation to normal coordinates is given by the linear canonical transformation

$$q_{j}=\sum\limits_{k=0}^{N-1}\sigma_{j,k}\xi_{k}\eqno(2)$$
with the coefficients
$$\sigma_{j,k}=\frac{1}{\sqrt{N}}\left(\sin\frac{2\pi kj}{N}+\cos\frac{2\pi kj}{N}\right),\ \  k = 0,1,\ldots, N-1.$$

The transformation $(2)$ allows us to present the quadratic part of Hamilton function as the total energy of the independent oscillators:
$$H_{2}=\sum\limits_{k=1}^{N-1}\frac{1}{2}(\eta_{k}^{2}+\omega_{k}^{2}\xi_{k}^{2}).\eqno(3)$$

Here $\xi_k$ and $\eta_k$ are the amplitudes and the momenta of NNMs; the coordinate $\xi_0$ associated with motion of the center of mass is removed from $(3)$. The eigenvalues $\omega_k$  have the form
$$\omega_{k}=\omega_{N/2}\sin\frac{\pi k}{N}, \ \ \omega_{N/2}=2, \ \ k=0,\ldots, N-1.\eqno(4)$$

If the number of particles increases, the frequency gap between the high\-est-frequency mode and nearby modes quickly decreases. In what follows, we consider an even number of particles $N$. In this case, the frequencies $\omega_k$ are bounded by the high\-est-fre\-quency $\omega_{N/2} = 2$. The zero eigenvalue $\omega_0 = 0$ corresponds to motion of the chain as a rigid body. All eigenvalues $(4)$ are in the interval $0 < \omega_k < 2$ and twice degenerate.

The remaining part of the Hamilton function $(1)$ for the periodic $(\alpha-\beta)$-FPU chain has the following form \cite{PoggiRuffo}:
$$H_{nonlin}=\frac{\alpha}{6\sqrt{N}}\sum \limits_{i,j,k=1}^{N-1}\omega_{i} \omega_{j} \omega_{k} D_{ijk} \xi_{i} \xi_{j} \xi_{k} +$$ $$+ \frac{\beta}{8N} \sum \limits_{k,l,m,n=1}^{N-1}\omega_{k} \omega_{l} \omega_{m} \omega_{n} \xi_{k} \xi_{l} \xi_{m} \xi_{n} C_{k l m n},\eqno(5)$$
where
$$D_{n,m,k} = -\Delta_{n+m+k}+\Delta_{n+m-k}+\Delta_{n-m+k}+\Delta_{n-m-k},$$
$$C_{k,l,m,n}= -\Delta_{k+l+m+n}+\Delta_{k+l-m-n}+\Delta_{k-l+m-n}+\Delta_{k-l-m+n}$$
and
$$
\left\{
\begin{array}{rcl}
            (-1)^{r},\  if\  r=mN, m\in \mathbb{Z} \\
             0\ \ \ \ \ \ \ \ \  \ \ \ \ \ \ otherwise \\
\end{array}
\right.
$$
So the equations of motion can be written as:
$$\frac{d^{2}\xi_{k}}{dt^{2}}+\omega_{k}^{2} \xi_{k}+\frac{\alpha}{2\sqrt{N}}\omega_{k}\sum \limits_{l,m=1}^{N-1}\omega_{l} \omega_{m} D_{lmk} \xi_{l} \xi_{m} + \ \ \ \ \ \ \ \ \ \ \ \ \ \ \ $$
$$\ \ \ \ \ \ \ \ \ \ \ \ \ \ \ +\frac{\beta}{2N}\omega_{k}\sum \limits_{l,m,n=1}^{N-1} \omega_{l} \omega_{m} \omega_{n} C_{klmn} \xi_{l} \xi_{m} \xi_{n}=0.\eqno (6)$$

To analyse the dynamics of the chain in the context of Eqs $(6)$, we introduce the complex amplitude corresponding to the combination of displacement and velocity in the new basis
$$\Psi_{k}=\frac{1}{\sqrt{2}}\left(\frac{\eta_{k}}{\lambda_{k}}+i\lambda_{k}\xi_{k}\right),\ \ \omega_{k}=:\lambda_{k}^{2},$$
$$\xi_{k}=-\frac{i}{\lambda_{k}\sqrt{2}}\left(\Psi_{k}-\Psi_{k}^{*}\right),\ \ \eta_{k}=\frac{\lambda_{k}}{\sqrt{2}}\left(\Psi_{k}+\Psi_{k}^{*}\right).$$
In terms of the complex amplitudes, Eqs $(6)$ can be rewritten as
$$\frac{d\Psi_{k}}{dt}-i\omega_{k} \Psi_{k}-\frac{\alpha}{4\sqrt{2N}} \sum \limits_{l,m=1}^{N-1} \lambda_{k} \lambda_{l} \lambda_{m} D_{lmk} \left(\Psi_{l}-\Psi_{l}^{*}\right) \left(\Psi_{m}-\Psi_{m}^{*}\right)+$$
$$ +\frac{\beta i}{2N}\sum \limits_{l,m,n=1}^{N-1} \lambda_{k} \lambda_{l} \lambda_{m} \lambda_{n} C_{klmn} \left(\Psi_{l}-\Psi_{l}^{*}\right) \left(\Psi_{m}-\Psi_{m}^{*}\right) \left(\Psi_{n}-\Psi_{n}^{*}\right)=0.\eqno{(7)}$$
Taking into account the dependence of the chain properties on the number of the particles, one can choose the value $1/\sqrt{N}$ as a small parameter $\varepsilon$.

\section{Main asymptotic approximations.}
We employ the multiple scale procedure \cite{ManMikhl} in order to consider the processes in the time intervals greatly exceeding $2\pi /\omega_{k}$. Following this procedure, we introduce time scales:

$$\tau_{0}=t,\ \  \tau_{1}=\varepsilon\tau_{0},\ \  \tau_{2}=\varepsilon^{2}\tau_{0}, \ldots \eqno(8)$$
where the "fast" time $\tau_{0}$  corresponds to the initial time scale of the system, while the slow times $\tau_{1}$, $\tau_{2}$, etc. correspond to the slowly varying envelopes. The envelope functions $\varphi_{k}$ are defined by the relation
$$\Psi_{k}=\varphi_{k}e^{i\omega_{k}\tau_{0}}.$$

We construct the asymptotic representation of $\varphi_{k}$ in the form
$$\varphi_{k}=\chi_{k,1}+\varepsilon\chi_{k,2}+\varepsilon^{2}\chi_{k,3}+\ldots \eqno(9)$$

The multiple scale expansion based on the relations $(8)$, $(9)$ will be used to derive the equations of the leading-order approximation $\chi_{k,1}$.

After some calculations one can get the equations for the pair of modes with the same eigenvalue (we denote $\chi_{k,1} =: \chi_{k}$ and suppose that the other modes are not excited.):
 
$$\frac{i \partial \chi_{k}}{\partial \tau_{2}} + c_{1,k} |\chi_{k}|^{2} \chi_{k} + c_{2, k} |\chi_{N - k}|^{2} \chi_{k} + c_{3,k} \chi_{N - k}^{2} \chi_{k}^{*} = 0\ \ \ \ \ \ \ \ \ \ $$
$$\frac{i \partial \chi_{N - k}}{\partial \tau_{2}} + c_{1,k} |\chi_{N - k}|^{2} \chi_{N - k} + c_{2, k} |\chi_{k}|^{2} \chi_{N - k} + c_{3,k} \chi_{k}^{2} \chi_{N - k}^{*} = 0,\eqno(10)$$
where
$$c_{1, k}=\left(\left(\frac{9 \beta}{8} - \frac{3 \alpha^{2}}{4}\right)\omega_{k}^{2} + \alpha^{2}\right),\ c_{2, k}=\left(\left(\frac{3 \beta}{4} + \frac{\alpha^{2}}{2}\right)\omega_{k}^{2} - 2 \alpha^{2}\right),$$
$$c_{3,k}=\left(\left(\frac{3 \beta}{8} - \frac{5 \alpha^{2}}{4}\right)\omega_{k}^{2} + 3 \alpha^{2}\right).$$

This system pos\-ses\-ses in\-teg\-rals $$X=|\chi_{N-k}|^{2} + |\chi_{k}|^{2},$$ and $$G_{-}=i (\chi_{k} \chi_{N-k}^{*} - \chi_{N-k} \chi_{k}^{*}),$$ so it can be rewritten as:
$$\frac{\partial \chi_{k}}{\partial \tau_{2}} + c_{3, k} G_{-} \chi_{N-k} - i c_{1, k} X \chi_{k} = 0$$
$$\frac{\partial \chi_{N-k}}{\partial \tau_{2}} - c_{3, k} G_{-} \chi_{k} - i c_{1, k} X \chi_{N-k} = 0$$

The effective linearization of the equation of motion $(10)$ clarifies why the normal modes may preserve their individuality in the nonlinear case. The intensive energy exchange corresponds to small value of $G_{-}$. This process is described by a phase trajectory which is far from the equilibrium point, and results in a slow rearrangement of energy by periodic waves. The rate of energy exchange increases as 
the value of $G_{-}$ grows, but amount of the energy participating
in the exchange decreases (Fig.1).

\begin{center}
\includegraphics[width = 42mm, height = 40mm]{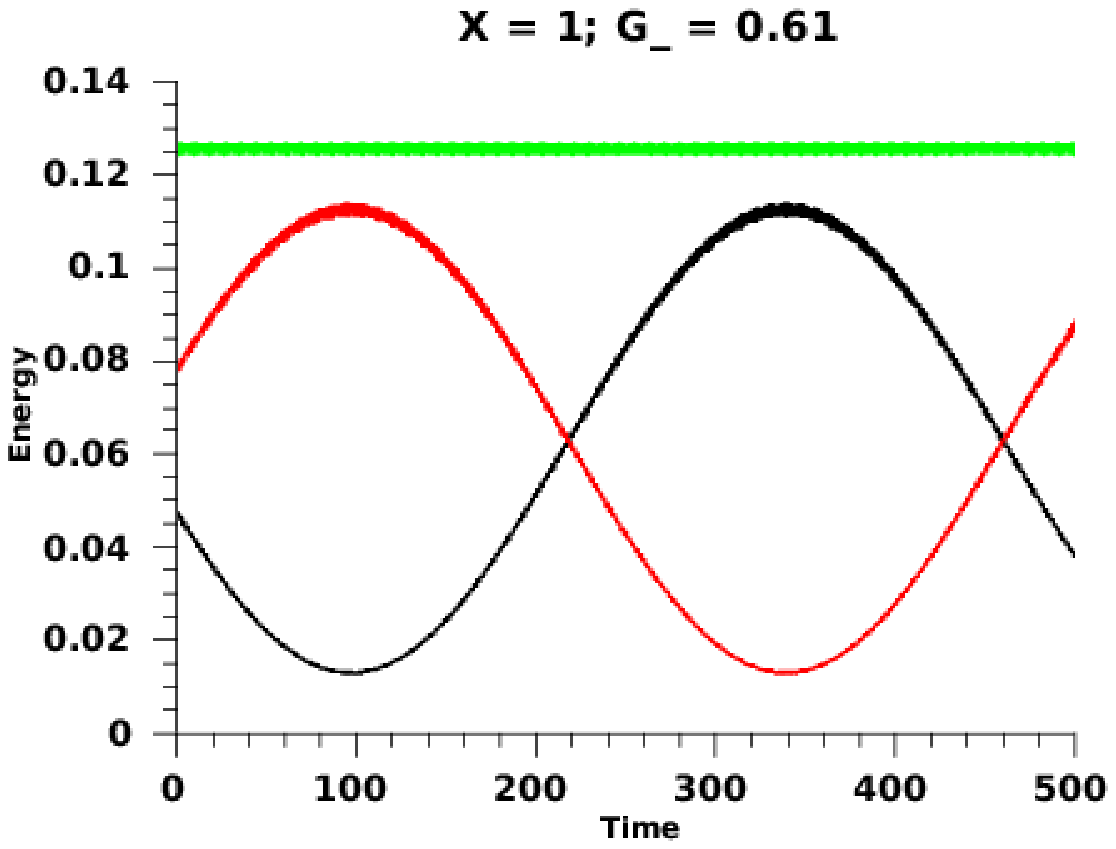}
\includegraphics[width = 44mm, height = 38mm]{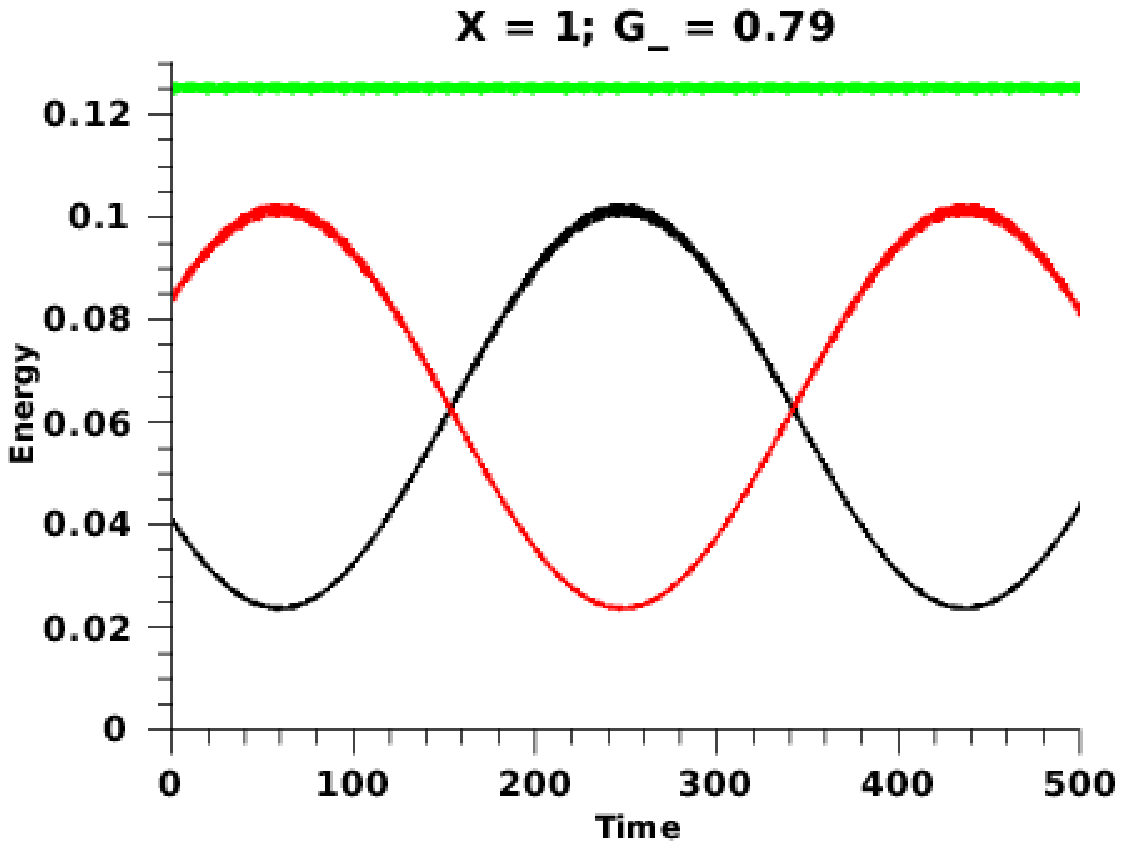}
\includegraphics[width = 44mm, height = 38mm]{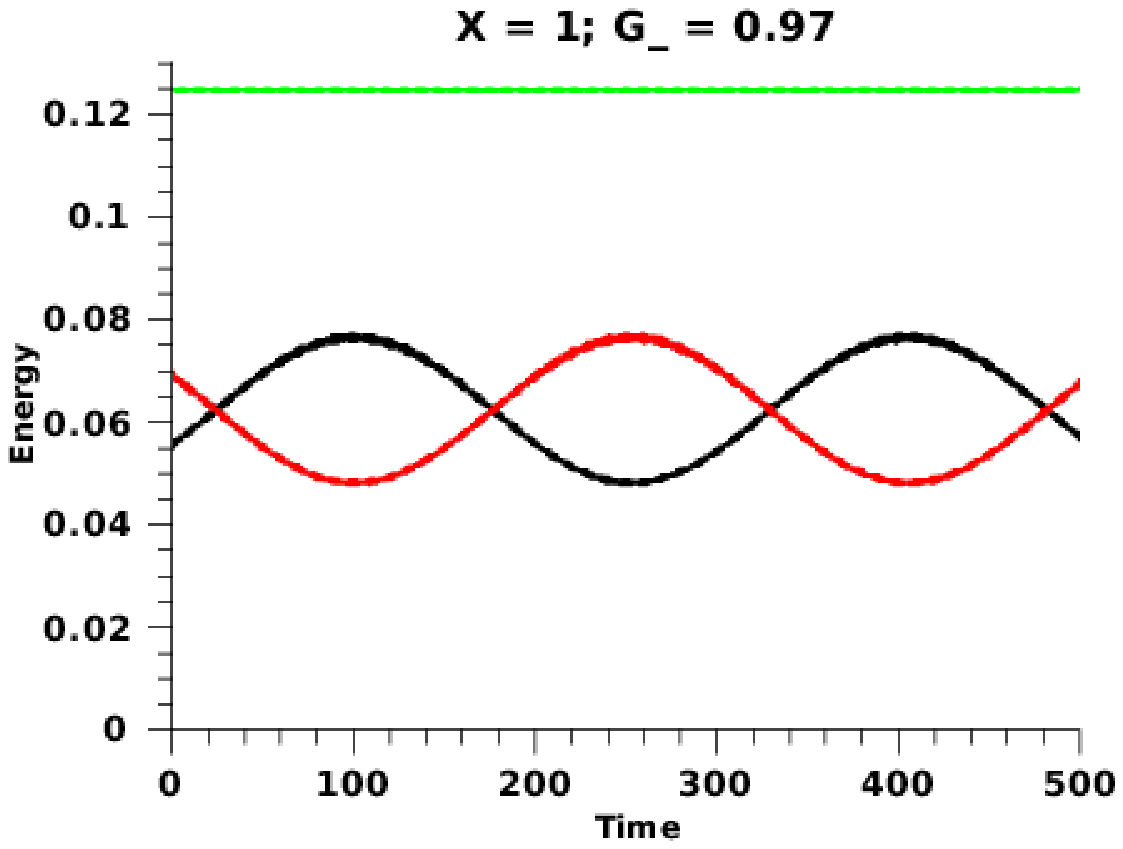}

Fig.1. The increase in energy exchange rate and the decrease of the energy 
exchange intencity while growing $G_{-}$.
\end{center}

The dynamics of the normal modes in the essentially discrete system (small value of $N$) should be considered in the assumption that the gap between $\pi$-mode and nearby conjugate ones is large enough. But, if the number of particles grows, then the new opportunity arises. Namely, the resonant interaction between the modes having close frequencies (in the linear approximation limit) becomes possible.

While the energy of excitation and/or number of particles grows, the dynamics of the chain becomes nonquasiharmonic. While the parameter $\varepsilon$ decreases, the eigenvalues of the $\pi$-mode and the nearby ones become more and more close until the resonant conditions occur.
$$\omega_{N/2-1} = \omega_{N/2}\left(1-\frac{\pi^{2}}{2 N^{2}}\right)$$

In such a case the equations (10) are not valid because this resonance is not taken into account. Now we choose the top frequency $\omega_{N/2}$ as the basic one. So, because of the closeness of frequencies, additional linear terms appear in the equations for $\chi_{N/2-1}$ and $\chi_{N/2+1}$. Moreover, we have to take into account the resonant interactions in the nonlinear part of equations.
The problem turns out to be more complicated than that for the $\beta$-FPU chain. After lengthy transformations (details of calculation one can see in Appendix) we have got the the equations
$$i \frac{\partial \chi_{1}}{\partial \tau_{2}} - \frac{\pi^{2}}{2} \Omega^{2} \chi_{1} + C \Omega^{2}\left(|\chi_{1}|^{2} \chi_{1} + \chi_{2}^{2} \chi_{1}^{*} + 2 \chi_{0}^{2} \chi_{1}^{*} + 2 |\chi_{0}|^{2} \chi_{1}\right) = 0$$
$$i \frac{\partial \chi_{2}}{\partial \tau_{2}} - \frac{\pi^{2}}{2} \Omega^{2} \chi_{2} + C \Omega^{2}\left(|\chi_{2}|^{2} \chi_{2} + \chi_{1}^{2} \chi_{2}^{*} + 2 \chi_{0}^{2} \chi_{2}^{*} + 2 |\chi_{0}|^{2} \chi_{2}\right) = 0$$
$$i \frac{\partial \chi_{0}}{\partial \tau_{2}} + 2 C \Omega^{2} \left((|\chi_{1}|^{2} + |\chi_{2}|^{2}) \chi_{0} + (\chi_{1}^{2} + \chi_{2}^{2}) \chi_{0}^{*})\right) = 0,\eqno(11)$$
where
$$C = \frac{3 \beta_{1}}{4} = \frac{3 \beta}{4} - \alpha^{2},\ \chi_{0} := \chi_{N/2}, 
\ \chi_{1} := \chi_{N/2-1},\ \chi_{2} := \chi_{N/2+1},\ \Omega := \omega_{N/2},$$
which coinside with those for the $\beta$-FPU chain if the constant of the
nonlinear interaction is renormalized ($\beta \leftrightarrow \beta_{1}$).  
This renormalization
is in a good agreement with the numerical results, e.g. the asymptotic system in
slow time obtained from FPU chain with such $\alpha$ and  $\beta$, that $\beta_{1} = 0$ 
behaves almost like linear one (Fig.2). So all effects described in \cite{PhysRevE}
also occur in $\alpha\beta$-FPU chain. Therefore, in the framework of this asymptotic 
approximation there is no difference between $\alpha\beta$- and $\beta$-FPU 
chains (the dynamics of $\alpha$-FPU system corresponding to $\beta < 0$). 
But, contrary to \cite{PhysRevE}, we do not restrict ourselves by numerical 
results and perform the analytical solution of $(11)$ for LPT. Besides, 
we discuss in more details the notion of the "effective particle".

\begin{center}
a)\includegraphics[width=60mm, height=35mm]{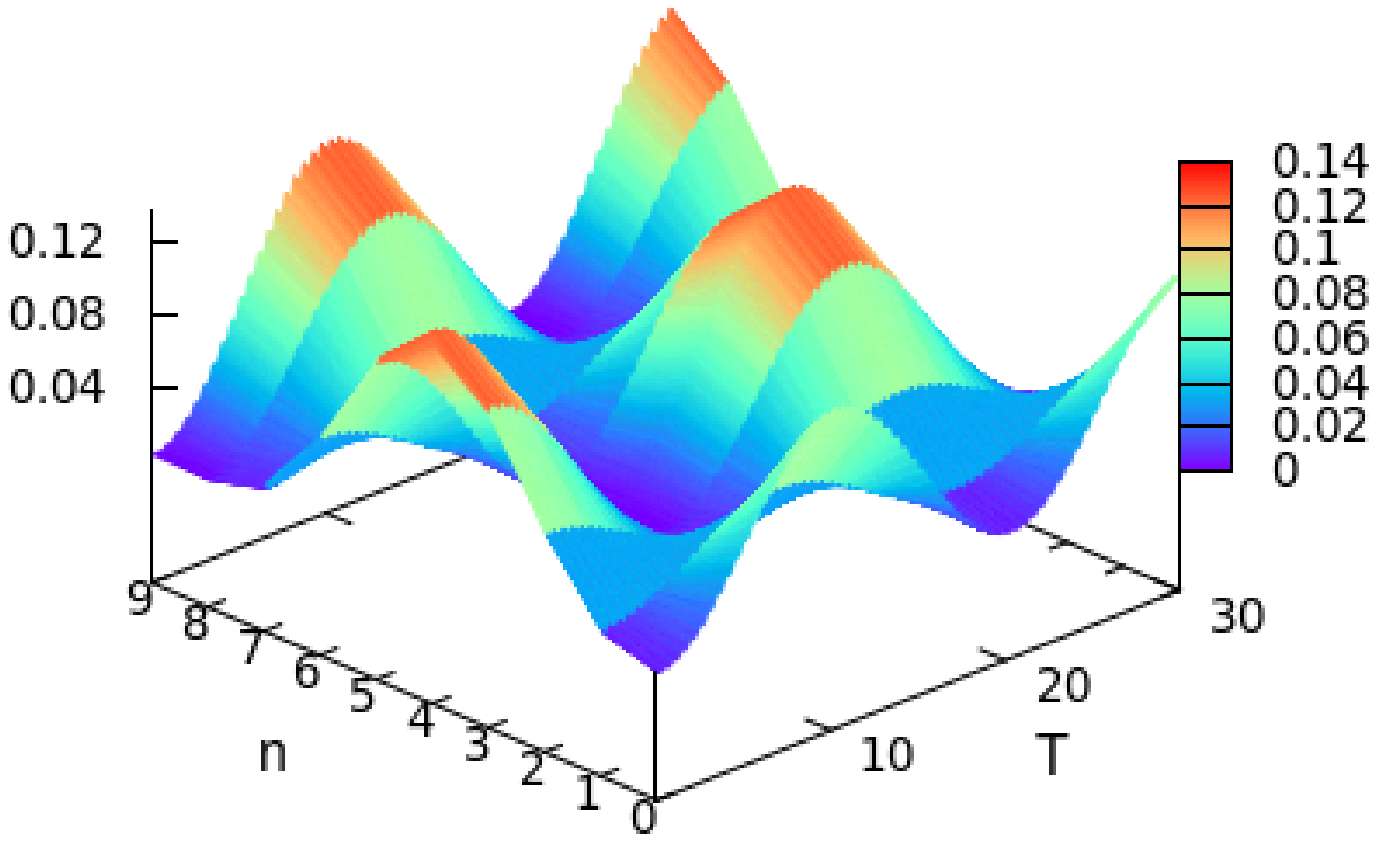} b)\includegraphics[width=60mm, height=35mm]{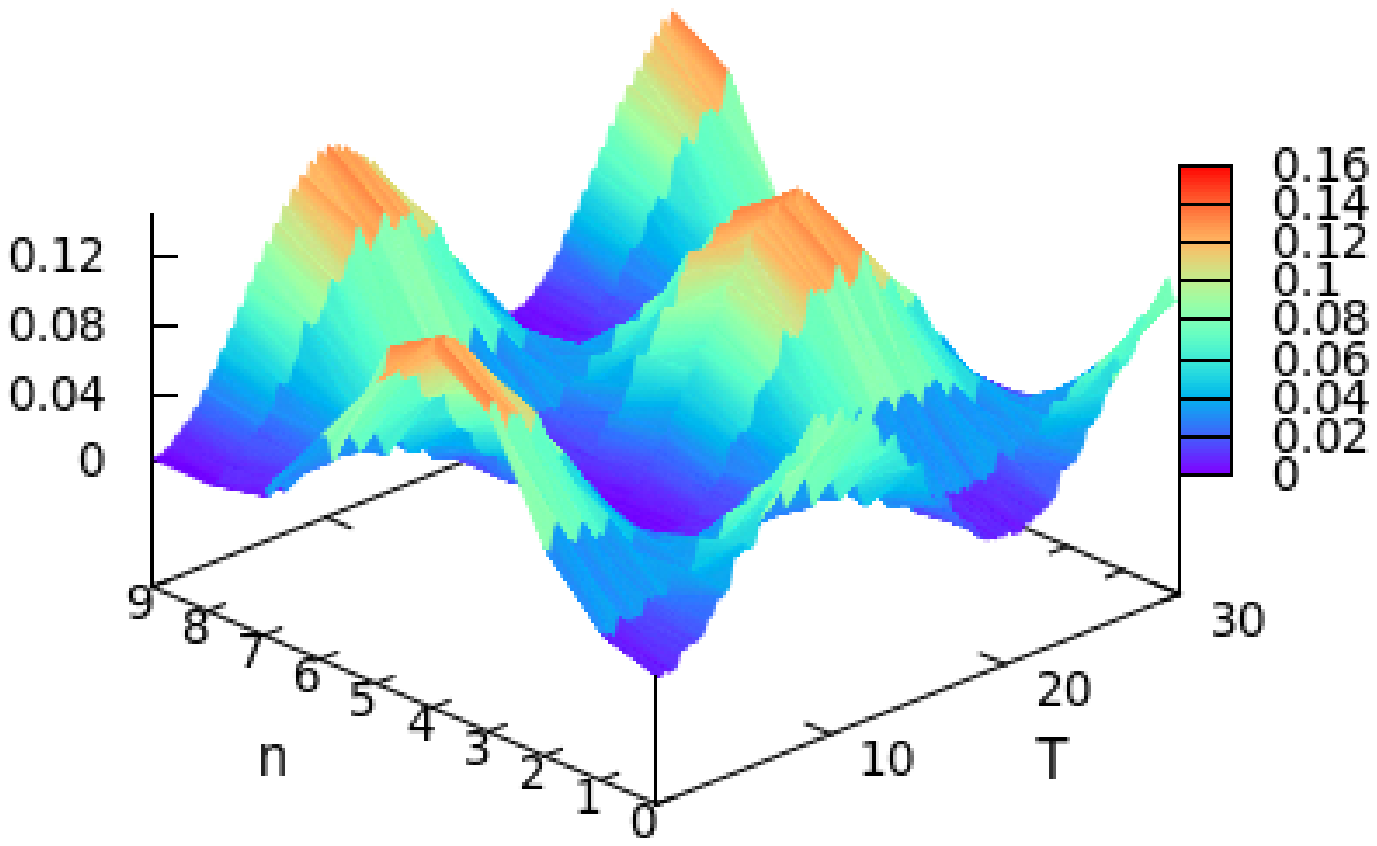}

Fig.2 Distribution of the energy among the particles of the chain; initial condition corresponds to LPT (excited one half of the chain -- "effective oscillator"): a) linear system ($\alpha = \beta = 0$) and b) "effectively linear" system ($\alpha = 0.5,\ \beta = 0.333$).
\end{center}

It is easy to check that equations possess an integral of motion -- total "occupation number":
$$X_{tot}:=|\chi_{1}|^{2} + |\chi_{2}|^{2} + |\chi_{0}|^{2}.$$

\section{From "waves" to "particles".}

The description of system's dynamics in the terms of NNMs has to be replaced by that in terms of the "effective particles". 
This allows us to describe the intensive energy exchange between different parts of the chain, which can be identified as "effective particles"\  with "coordinates"\  $\psi_{1}$ and $\psi_{2}$:
$$\psi_{1} = \frac{1}{\sqrt{2}}\left[\chi_{0} - \left(\chi_{1} \sqrt{1-c^{2}} + c \chi_{2}\right)\right],$$ 
$$\psi_{2} = \frac{1}{\sqrt{2}}\left[\chi_{0} + \left(\chi_{1} \sqrt{1-c^{2}} + c \chi_{2}\right)\right],$$ 
$$\varphi = \left(c \chi_{1} - \chi_{2} \sqrt{1-c^{2}}\right),$$
where $c$ is a constant defined by initial conditions $(0 \leq c \leq 1)$. Such transformation preserves the total occupation number in the form $X = |\psi_{1}|^{2} + |\psi_{2}|^{2} + |\varphi|^{2}$. For the value $c = \frac{1}{\sqrt{2}}$, which corresponds to equal initial conditions of modes, we have the equations in the coordinates of "effective particles":

$$i \frac{\partial \psi_{1}}{\partial \tau_{2}} + \frac{\pi^{2}}{4} \Omega^{2} (\psi_{2} - \psi_{1}) + C \Omega^{2}\left(|\psi_{1}|^{2} \left(\frac{9}{4} \psi_{1} - \frac{1}{2} \psi_{2} \right) - |\psi_{2}|^{2} \left(\frac{3}{2} \psi_{1} + \frac{1}{4} \psi_{2} \right)+\right.\ \ $$
$$\ \ \ \ \ \left. + \frac{1}{4}( \psi_{1}^{2} \psi_{2}^{*} - \psi_{2}^{2} \psi_{1}^{*} ) + |\varphi|^{2}( \psi_{1} + \psi_{2}) + \varphi^{2} \left(\frac{3}{2} \psi_{1}^{*} + \frac{1}{4} \psi_{2}^{*} \right)\right) = 0$$
$$i \frac{\partial \psi_{2}}{\partial \tau_{2}} + \frac{\pi^{2}}{4} \Omega^{2} (\psi_{1} - \psi_{2}) + C \Omega^{2}\left(|\psi_{2}|^{2} \left(\frac{9}{4} \psi_{2} - \frac{1}{2} \psi_{1} \right) - |\psi_{1}|^{2} \left(\frac{3}{2} \psi_{2} + \frac{1}{4} \psi_{1} \right)+\right.\ \ $$
$$\ \ \ \ \ \left. + \frac{1}{4}( \psi_{2}^{2} \psi_{1}^{*} - \psi_{1}^{2} \psi_{2}^{*} ) + |\varphi|^{2}( \psi_{2} + \psi_{1}) + \varphi^{2} \left(\frac{3}{2} \psi_{2}^{*} + \frac{1}{4} \psi_{1}^{*} \right)\right) = 0$$
$$i \frac{\partial \varphi}{\partial \tau_{2}} - \frac{\pi^{2}}{2} \Omega^{2} \varphi + C \Omega^{2} \left(\frac{}{}\varphi\left((|\psi_1|^{2} + |\psi_{2}|^{2} + |\varphi|^{2}) +(\psi_{1} \psi_{2}^{*} + \psi_{2} \psi_{1}^{*})\right)\right. +\ \ \ \ \ \ \ \ \ $$
$$\left. + \varphi^{*} \left(\frac{3}{2}(\psi_{2}^{2} + \psi_{2}^{2}) + \psi_{1} \psi_{2}\right)\right) = 0.\eqno(12)$$ 

Since the notion of effective particle is of great importance, we discuss it in more details. When dealing with the mechanisms of
intensive energy exchange, one needs to reveal the
elementary agents which exchange energy and become the domains of its
localization (after exceeding  some critical level of excitation). In
gaseous media, they are the interacting particles (atoms or molecules)
themselves, which participate in  almost free motion. 

On the contrary,
the particles in oscillatory chains as well as in all crystalline
solids undergo a strong mutual interaction. Therefore, the elementary
agent here is all oscillatory chain performing oscillations
corresponding to one of non-interacting normal modes. However, an
increase of the particles number leads to the resonance relations between
several Normal Modes and then between multiple modes with close frequencies.
If the initial conditions are strongly asymmetric (they are far from those
corresponding to every of resonating  normal  modes), it leads to the
manifestation of two significant effects. 

First of  them is the inter-modal
coherence and second one is the inter-modal interaction (in nonlinear
case). As a result, the resonating  normal modes are no longer the
appropriate elementary agents. In the system of two weakly coupled
oscillators their role is played by the particles themselves (as well as in
the case of gaseous media) changing slowly by the energy in the
beating process (their displacements can be presented as a sum and a
difference of the modal variables). 

As the number of particles increase, the appearance of the resonating modes does
not mean that we have to come back to the real particles. Now the "effective 
particles" play a role of elementary agents.

Their displacements can be constructed as combinations of resonating modal
variables only (similarly to the displacements of two weakly coupled
oscillators) which manifest the inter-modal coherence in the beating
process described by limiting phase trajectories. Thereby, the beating
notion is extended to multidimensional systems. Besides, the
introducing the LPT allows us to describe adequately the transition from
intensive energy exchange to energy localization due to an increase of
intensity of excitation. By this manner the connection with continuum
systems, having the breathers as localized solutions is clarified.
At last, a most simple analytic presentation of intensive energy
exchange can be  attained precisely in terms of the "effective particles" and
Limiting Phase Trajectories.

Let us consider a particular solution corresponding to $\varphi = 0, \ \chi_{1}  = \chi_{2}$. In this case, the Hamilton function is
$$H = \Omega^{2} \left(-\frac{\pi^{2}}{4} |\psi_{1} - \psi_{2}|^{2} + C \left(9 (|\psi_{1}|^{2} + |\psi_{2}|^{2})^{2} + 6 |\psi_{1}|^{2} |\psi_{2}|^{2} -\right.\right.$$$$\left.\left. - 2 (|\psi_{1}|^{2} + |\psi_{2}|^{2}) (\psi_{1} \psi_{2}^{*} + \psi_{2} \psi_{1}^{*}) + \psi_{1}^{2} (\psi_{2}^{*})^{2} + \psi_{2}^{2} (\psi_{1}^{*})^{2}\right)\frac{}{}\right).\eqno(13)$$

Since the occupation number $X = |\psi_{1}|^{2} + |\psi_{2}|^{2}$ is the integral of motion, the variables $\psi_{1}$ and $\psi_{2}$ can be expressed through the angle variables $\theta, \delta_{1}, \delta_{2}$:
$$\psi_{1} = X \cos{\theta} e^{i \delta_{1}},\ \ \psi_{2} = X \sin{\theta} e^{i \delta_{2}}\eqno(14)$$
Substitution of $(14)$ with $\varphi = 0$ into $(12)$ yields the following equations 
$$\frac{d \theta}{d \tau_{2}} + K_{1} \sin{\Delta} + K_{2} \sin{2 \theta} \sin{2 \Delta} = 0\ \ \ \ \ \ \ \ \ \ \ \ \ \ \ \ \ \ \ \ \ \ \ \ \ \ \ \ \ \ \ \ \ \ \ \ $$
$$\frac{d \Delta}{d \tau_{2}} \sin{2 \theta} + 2 K_{1} \cos{\Delta} \cos{2 \theta} - 4 K_{2} \sin{2 \theta} \cos{2 \theta} \left(8 - \cos^{2}{\Delta}\right) = 0\eqno{(15)}$$
where
$$\Delta = \delta_{1} - \delta_{2}, \ K_{1} = \Omega^{2} \left(\frac{\pi^{2}}{4} - \frac{3 \beta_{1} X}{32}\right); \ K_{2} = \frac{3 \beta_{1} X}{64} \Omega^{2};\ \beta_{1} = \beta - \frac{4}{3} \alpha^{2} = \frac{4}{3} C.$$

Note that Eqs $(15)$ conserve the first integral 
$$H(\theta, \Delta) = X \left(\frac{27 \beta_{1} X - 16 \pi^{2}}{64} \Omega^{2} + K_{1} \sin{2 \theta} \cos{\Delta} -\right.$$
$$\left.- K_{2}\frac{}{} \left(8 - \cos^{2}{\Delta}\right) \sin^{2}{2 \theta}\right)\eqno{(16)}$$
obtained from $(13)$ by changing $\psi_{1},\psi_{2}$ to expressions $(14)$.

Equations (16) for all $X, \ \beta_{1}$ have the following stationary points (Fig.3):
$$(a)\ \Delta = 0,\ \theta = \frac{\pi}{4};\ \ (b)\ \Delta = \pi,\ \theta = \frac{\pi}{4}\eqno(17)$$
The points $(a)$ and $(b)$ correspond to $\psi_{1} = \psi_{2}$; by definition, this leads either to the pure $\pi$-mode $(a)$ or to the pure $(\chi_{N/2-1} + \chi_{N/2+1}))$-mode $(b)$.

\begin{center}
\includegraphics[height=70mm]{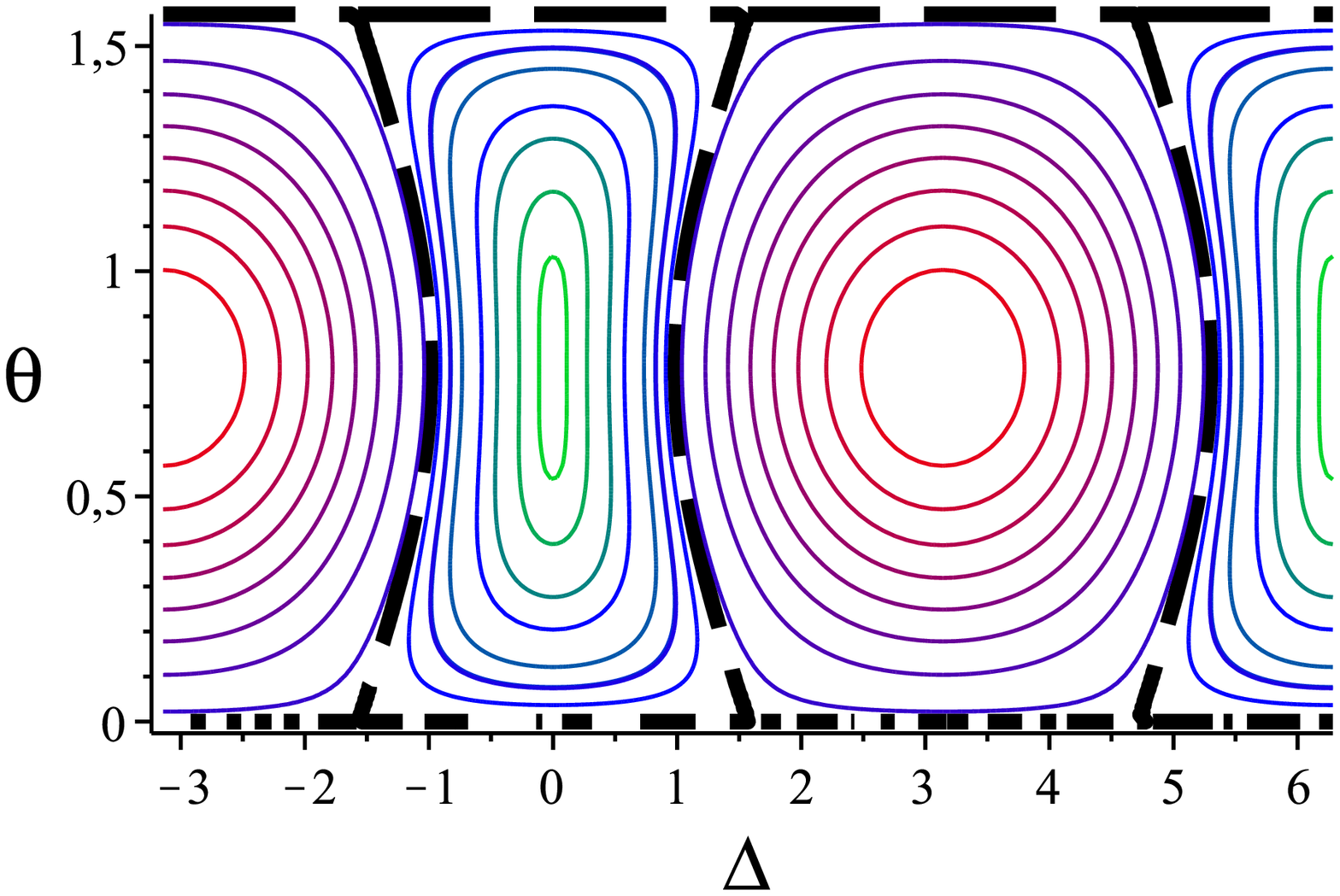} 

Fig.3 Phase portrait of Eqns $(15)$ with $X <X_{c}$ (LPT -- dashed line) .
\end{center}

When the amplitude of excitation grows, the in-phase stationary point $(17a)$ becomes unstable (Fig.4). This instability leads to the appearance of two new stationary points. Namely, if the occupation number $X$ exceeds the value $X_{c} = \frac{\pi^{2}}{3 \beta_{1}}$ then there exist three stationary solutions of Eqns $(15)$ with the phase shift $\Delta = 0$:
$$(a)\ \ \Delta = 0,\ \theta = \frac{\pi}{4};$$
$$(b)\ \ \Delta = 0,\ \theta = \frac{1}{2} \arcsin{\frac{K_{1}}{14 K_{2}}}\eqno(18)$$
$$(c)\ \ \Delta = 0,\ \theta = \frac{\pi}{2} - \frac{1}{2} \arcsin{\frac{K_{1}}{14 K_{2}}}$$

\begin{center}
a)\includegraphics[width = 60mm,height=40mm]{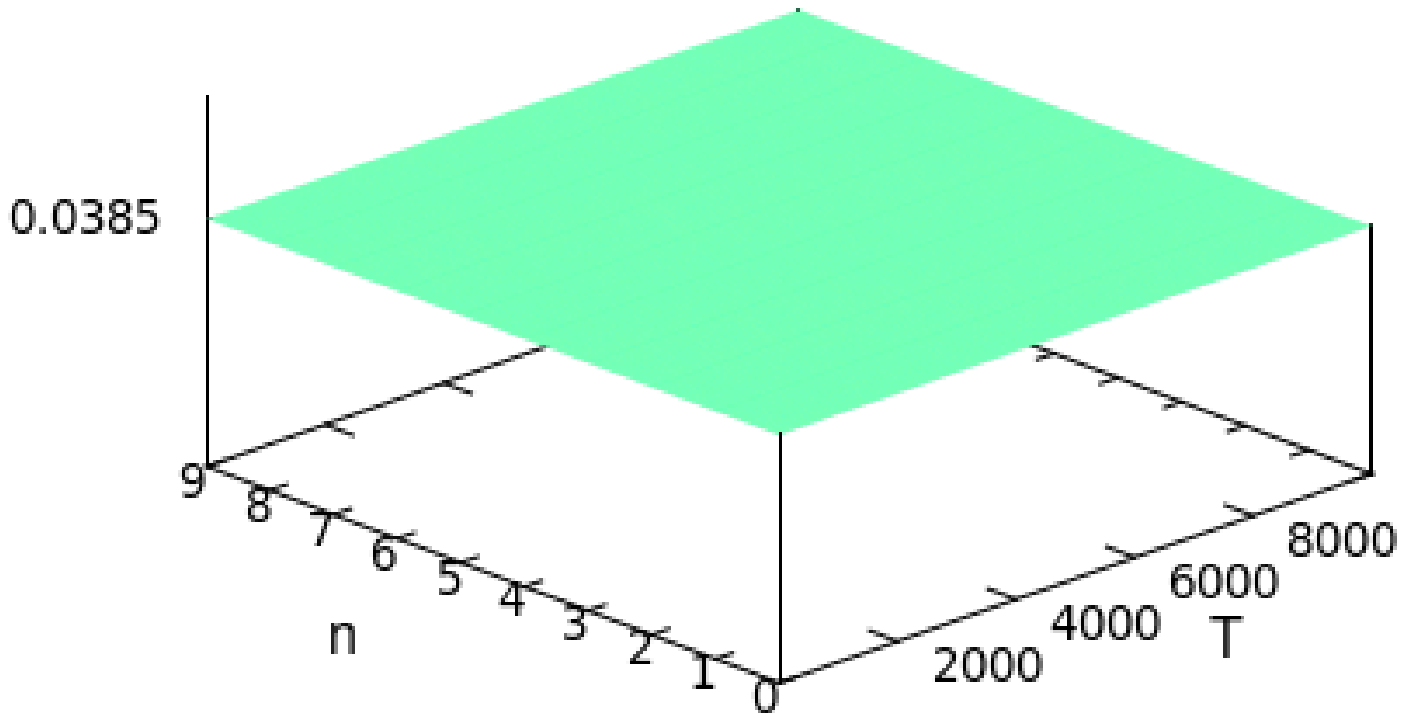} b)\includegraphics[width = 60mm,height=40mm]{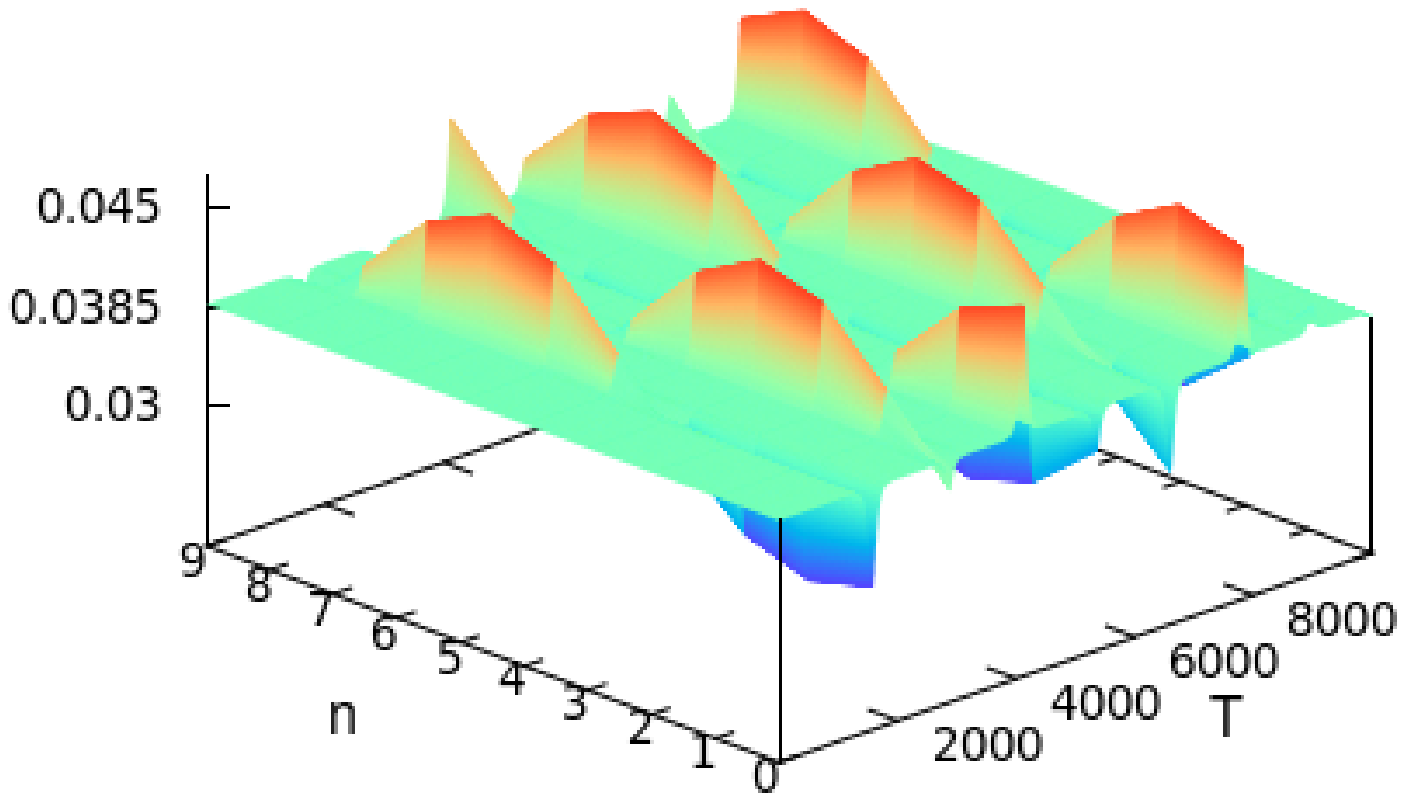}

Fig.4 Distribution of the energy among the particles of the chain (numerical solution); initial condition corresponds to $\pi$-mode: a) before the  treshold of instability $X_{c}$ and b) after it.
\end{center}

The threshold of instability of the $\pi$-mode is equal to $X_{c} = \pi^{2} / 3 \beta_{1}$, which is in a good accordance with the estimation obtained in the framework of the "narrow packet"\  approximation \cite{FPU}. Any trajectories in the neighborhood of stationary points $(18b, 18c)$ correspond to weakly localized solutions, for which the energy of one half of the chain -- "effective particle"\ -- only slightly exceeds the energy of the second one. At the same time, the trajectories starting beyond this vicinity and passing through the points $\psi_{1}\ (\theta = 0)$ or $\psi_{2}\ (\theta = \pi / 2)$ correspond to the combinations of modes with approximately equal energies that means that an initial excitation of one "effective particle"\ entails the complete energy exchange between both ones, i.e., the transition from the state $\psi_{1}$ to the state  $\psi_{2}$ and inversely. This implies that a possibility of complete energy exchange between different parts of the chain exists for the excitation level exceeding the instability threshold $X_{c}$ (Fig.5).

\begin{center}
a)\includegraphics[width = 55mm,height=40mm]{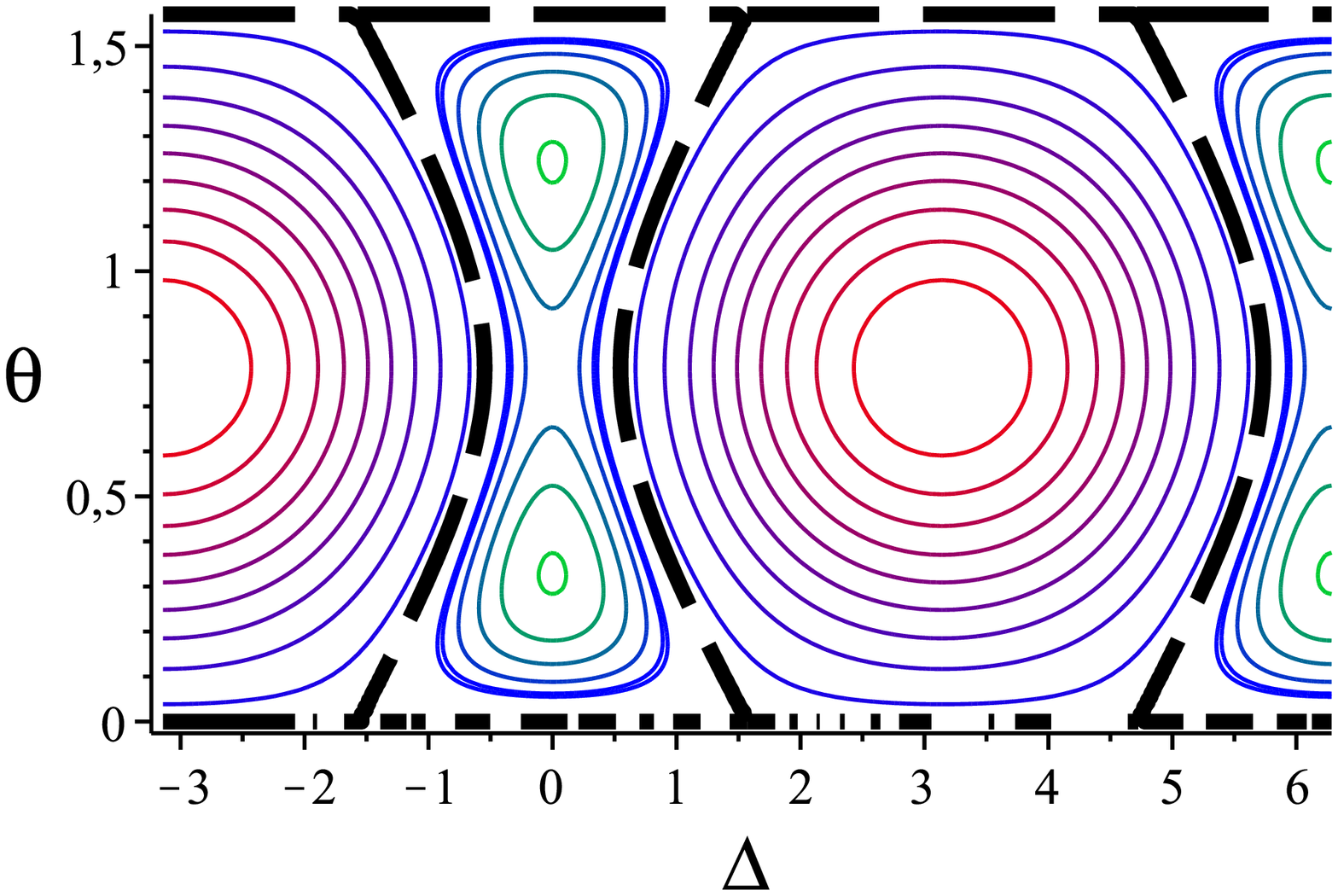} b)\includegraphics[width=70mm,height=40mm]{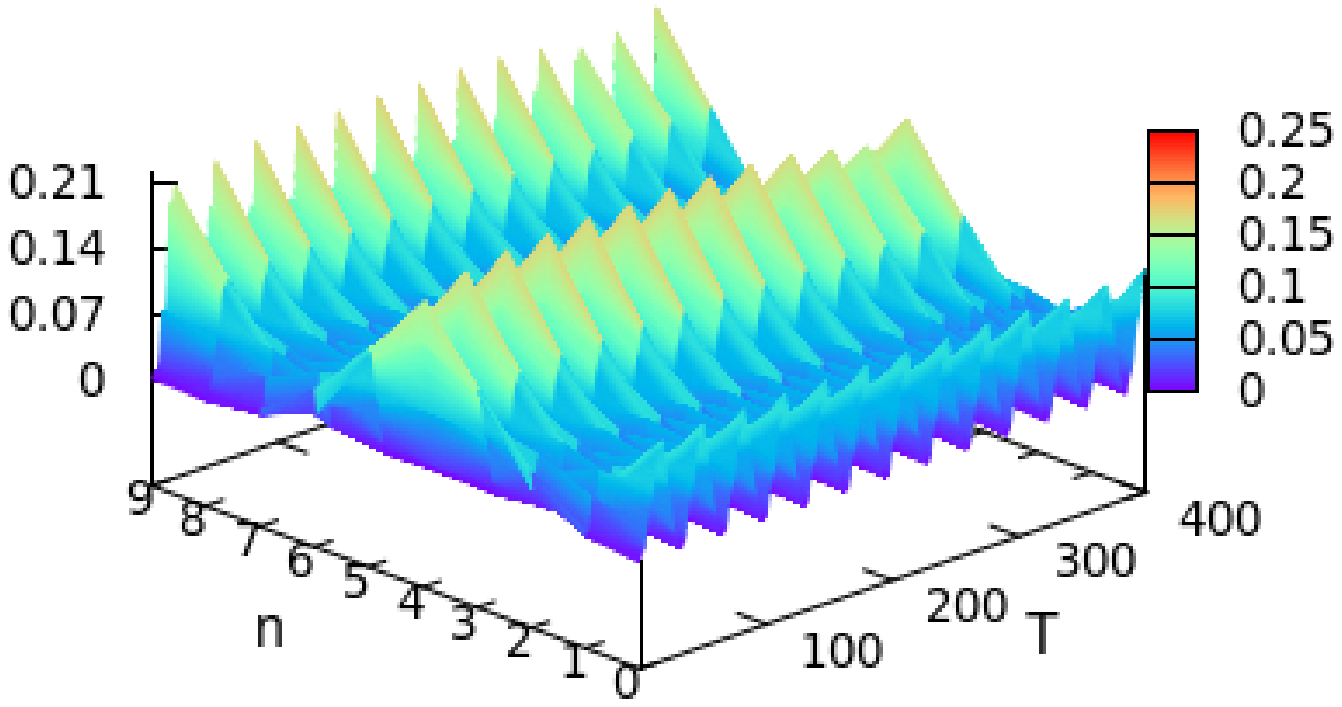}

Fig.5 a) Phase portrait of Eqs $(15)$ (LPT -- dashed line) and b) distribution of energy among the particles in the chain along LPT (numerical solution), $X_{c} < X < X_{loc}$, .
\end{center}

A growth of the amplitude $X$ entails an enlargement of the domain encircled by the separatrix passing through an unstable stationary point (out-of-phase NNM); at last, the separatrix coincides with the LPT. At this instant, the topology of the phase plane changes drastically (Fig. 5). Namely, any energy exchange between the mixed states $\psi_{1}$ and $\psi_{2}$ disappears; this implies that a trajectory starting at a point corresponding to $\theta < \pi / 4$ (or $\theta > \pi / 4$) and for any $\Delta$ cannot reach a point corresponding to $\theta > \pi / 4$ (or $\theta < \pi / 4$) [excepting the trajectories surrounding in-phase stationary points $\theta = \pm \pi$ and bounded by the separatrix going through the unstable point $\theta < \pi / 4,\ \Delta = 0$]. Therefore, the energy initially concentrated near the states $\psi_1$ or  $\psi_2$ remains confined in the excited effective oscillator (Fig. 6).

\begin{center}
a)\includegraphics[width = 55mm,height=40mm]{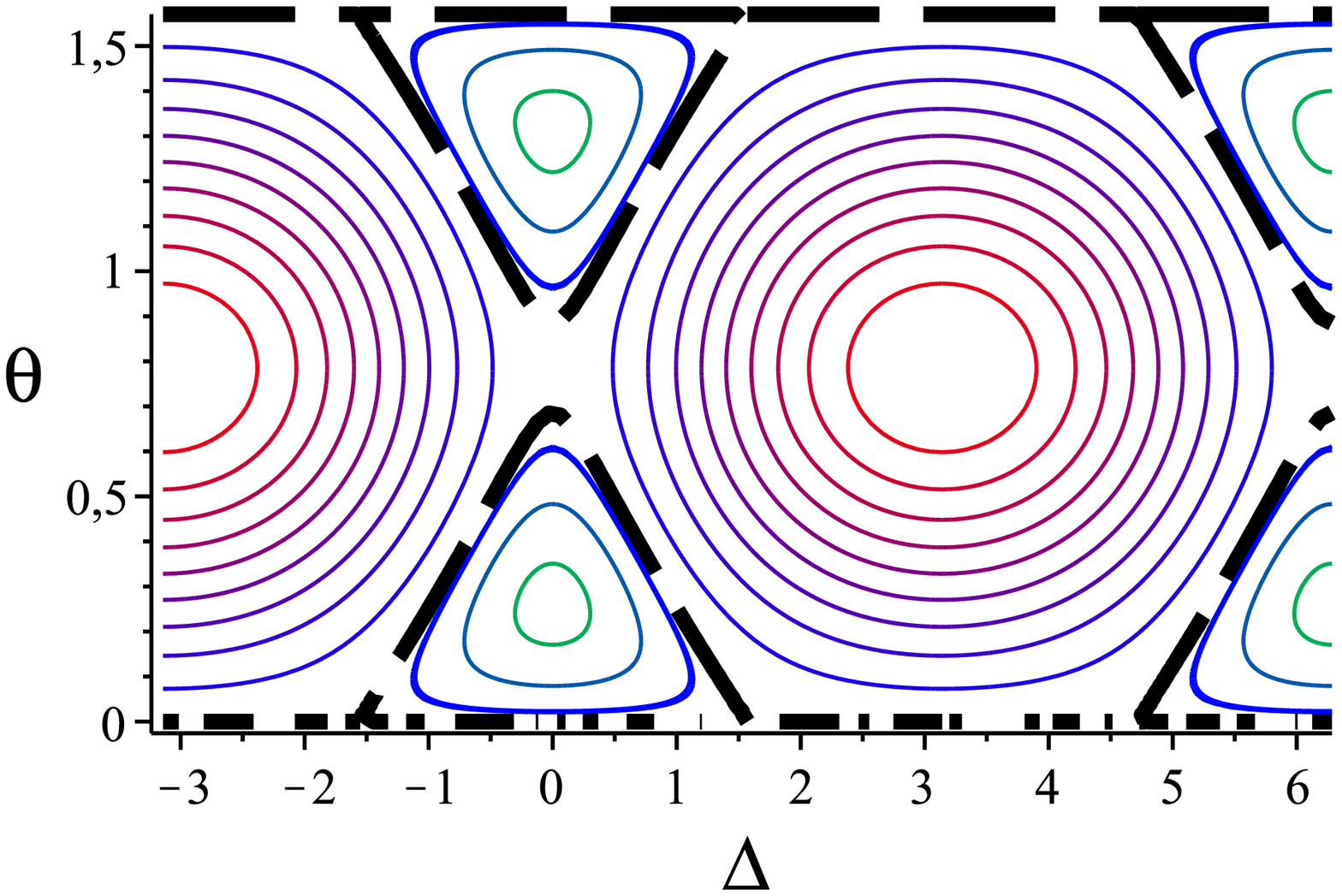} b)\includegraphics[width=70mm,height=35mm]{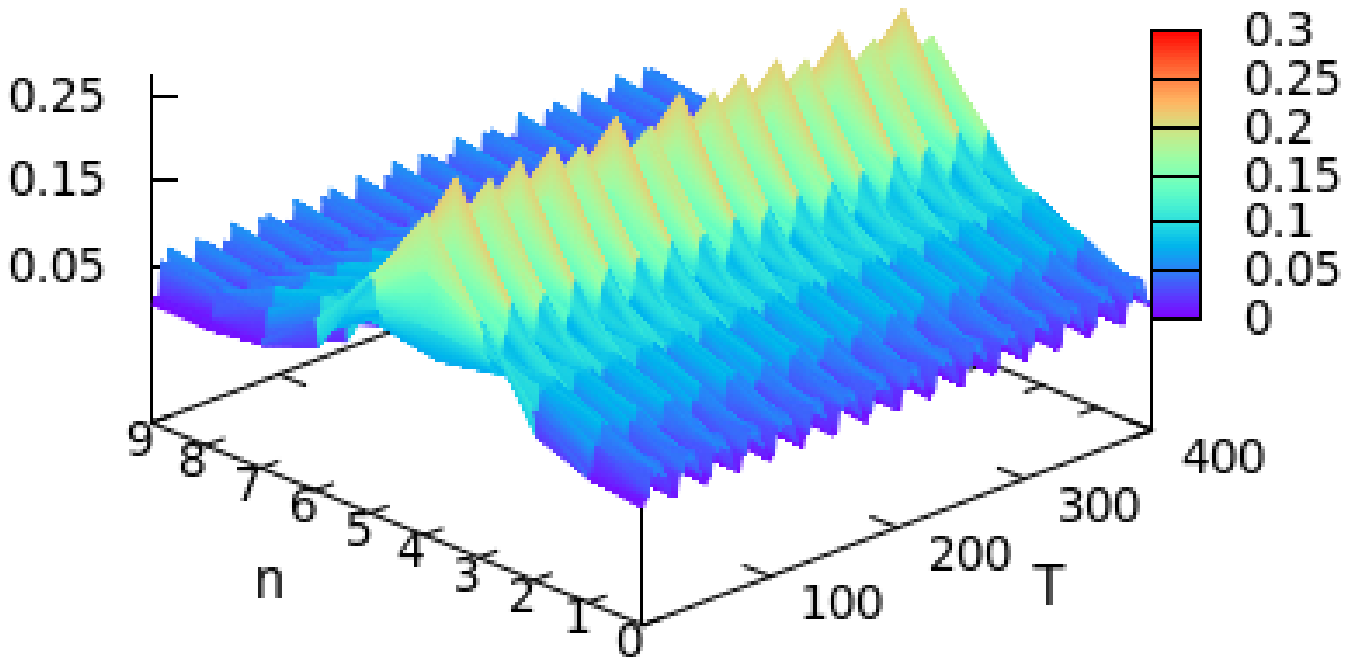}

Fig.6 a) Phase portrait of Eqns $(16)$ (LPT -- dashed line) and b) distribution of energy among the particles in the chain along LPT (numerical solution), $X > X_{loc}$ .
\end{center}

An energy threshold associated with the above mentioned localization can be found from the condition of equality of the energy corresponding to the LPT and the energy at the unstable $\pi$-mode. It is seen in Figs. 3, 5, 6 that the LPT goes through the points $\theta = \pi / 2$ and $\theta = 0$. This means that
$$H(\theta, \Delta)|_{LPT} = \frac{\Omega^{2} X}{64}\left(27 \beta_{1} X - 16 \pi^{2}\right)$$
$$\cos{\Delta}=\frac{K_{2}}{K_{1}} \sin{2 \theta}\left(8 - \cos^{2}{\Delta}\right),$$
$$K_{1} = \Omega^{2} \left(\frac{\pi^{2}}{4} - \frac{3 \beta_{1} X}{32}\right);\ K_{2} = \frac{3 \beta_{1} X}{64} \Omega^{2}.$$ 

Since the energy of the $\pi$-mode is equal to zero, it is easy to calculate the respective occupation number $X_{loc} = 16 \pi^{2} / 27 \beta_{1}$ and the energy of the chain $E_{loc} = 16 \pi^{2} / 27 \beta_{1} N$. It now follows that above the excitation level $E_{loc}$ we can observe the localized vibration excitation (a breather).

If the initial energy is concentrated at the state $\psi_{1}$ or $\psi_{2}$, the representing point in the phase plane moves along a trajectory encircling the respective stationary point. Then the temporal evolution of the breather corresponds to the regular variation of its profile (the "breathing" mode of the localized excitation). The period of breathing can be calculated as
$$T = \oint{dt} = \frac{1}{\varepsilon^{2}} \oint{\frac{d\Delta}{d\Delta / d\tau_{2}}},$$
where the last integral is taking along the LPT.

Contrary to "breathing" breathers corresponding to the motion along the LPT, the
stationary points $(19)$ determine new normal modes with invariable non-homogeneous energy
distribution, or the breather-like excitations. In spite of these modes existence at any $X>X_{c}$,
the true threshold of localization is equal to $X_{loc}$ because the possibility of the complete
energy exchange is preserved for motion along the LPT till $X=X_{loc}$.

Thus we get the solution $(18)$ corresponding to immobile breather. Finally, we want to
clarify the nature of travelling breathers. We recall that the above results have been obtained
under assumption $\varphi=0$ that corresponds to equal amplitudes of the conjugate modes. Now we
assume that the amplitudes of the conjugate modes are slightly different, so we consider
small enough $\varphi\neq 0$. One can show that the equations for functions $\psi_{1}$ and $\psi_{2}$ include only
quadratic terms depending on $\varphi$. Therefore, in the framework of the linear approximation, if
the value $\varphi\neq 0$ is small enough, there is no qualitatively change in the phase-plane portrait in
Figs (3, 5a, 6a). As for the behaviour of $\varphi$ in the vicinity of any of the stationary points $(18)$, it is
described as follows:
$$i\frac{d\varphi}{d\tau_{2}}-\frac{\pi^{2}}{2}\varphi+\frac{3\beta_{1}X}{8}(2\varphi\sin2\theta+(3+e^{2i\delta_{1}}sin2\theta)\varphi^{*})=0.$$

The respective eigenvalue
$$\lambda=\pm\frac{\sqrt{891\beta_{1}^{2}X^{2} + 60\pi^{2}\beta_{1}X - 96\pi^{4}}}{28}$$
is imaginary if $X<X_{loc}$, i.e. if the amplitude of excitation is less than the threshold
corresponding to coincidence of LPT and separatrix.

\begin{center}
\includegraphics[width=40mm, height=80mm]{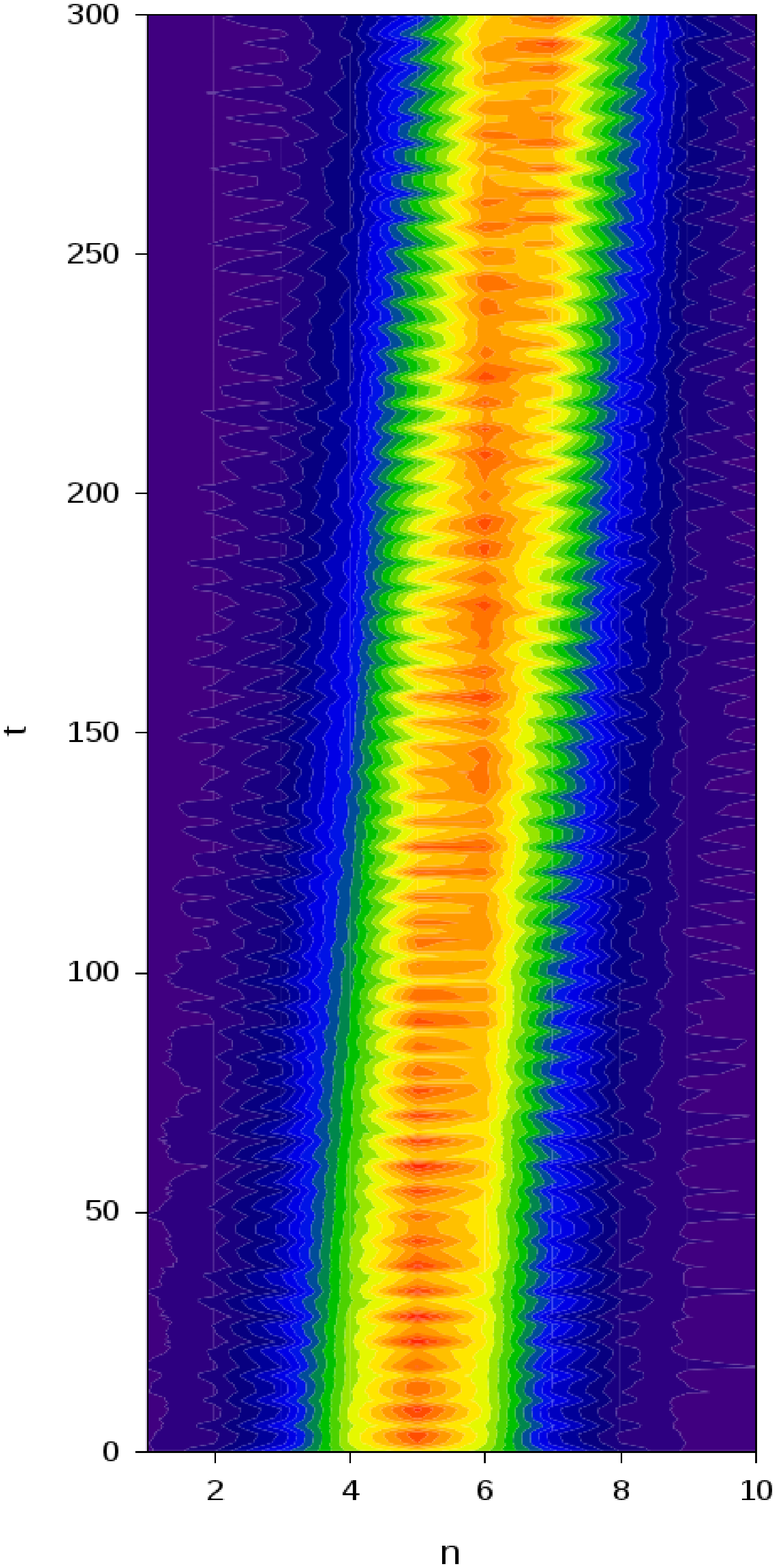} 

Fig.7 Directional motion of the breather when a real part in the eigenvalue $\lambda$ at $X>X_{loc}$.
\end{center}

It is easy to demonstrate that small non-zero $\varphi$ in the equations of motion leads to small oscillations of the breather center when the eigenvalue $\lambda$ is imaginary. From the other hand, appearance of a real part in the eigenvalue $\lambda$ at $X>X_{loc}$ leads to the directional motion of the breather (Fig. 7).

\section{Analytical solution for the LPTs.}
Using the equation for LPT
$$\cos{\Delta}=\frac{K_{2}}{K_{1}} \sin{2 \theta}\left(8 - \cos^{2}{\Delta}\right),$$
one can get the solution for LPT in terms of elliptical integrals:
$$\frac{d \Delta}{d \tau_{2}} - 2 \sqrt{K_{2}^{2} \left(8 - \cos^{2}{\Delta}\right)^{2} - K_{1}^{2} \cos^{2}{\Delta}} = 0,$$
$$\tau_{2} = \int\limits_{-\frac{\pi}{2}}^{\Delta}{\frac{d \Delta_{*}}{2\sqrt{K_{2}^{2} \left(8 - \cos^{2}{\Delta_{*}}\right)^{2} - K_{1}^{2} \cos^{2}{\Delta_{*}}}}}$$
$$\tau_{2} = \int\limits_{0}^{cos^{2}{\Delta}}{\frac{dz}{4 \sqrt{(z - z^{2})(K_{2}^{2} (8 - z)^{2} - K_{1}^{2} z)}}}.$$

We get the formal solution of $(15)$ to LPT, but, since its use is prohibitively difficult, 
we analyse the system in angle variables $(15)$ with LPT equation with $X < X_{loc}, X > X_{loc}$ and derive an
asymptotic formula to a periodic solution of these equations.

It is seen in Fig.8 that $\theta(\tau_{2})\ (X < X_{loc})$ and $\Delta(\tau_{2})\ (X > X_{loc})$ forms are close to the straight line, with an almost instant reverse at $\pi/2$ in the case of $\theta$ and instant step at $-\pi/2$ in the case of $\Delta$. In order to describe the solution we introduce the new variable $\tau(\phi), \phi = \Omega_{*}\tau_{2}$, $\Omega_{*}$ -- constant, which will be yielded from the first order approximation. The function $\tau(\phi)$ and its derivative $e(\phi)=d\tau/d\phi$ have the form

$$\tau(\phi)=\frac{2}{\pi}|\arcsin(\sin\frac{\pi\phi}{2})|,$$
$$\ \ e(\phi)=1,\ 0<\phi<1;\ e(\phi)=-1,\ 1<\phi<2.\eqno(19)$$

Following \cite{Vakakis}, we construct an approximate solution as a function of $\tau$; the inverse
transformation $\tau_{2}=\tau_{2}(\tau)$ automatically yields a periodic solution in $\tau_{2}$.
Taking into account the discontinuity of $\Delta(\tau_{2})$, we construct the solution of $(15)$ in the
form
$$\theta(\tau_{2})=\Theta(\tau),\ \ \Delta(\tau_{2})=e(\phi)Y(\tau), \ \ \frac{d}{d\tau_{2}}=\Omega_{*} e\frac{d}{d\tau}\eqno(20)$$

To derive the equations for $\Theta(\tau)$, $Y(\tau)$, we insert $(19)$ into $(15)$. This yields the set of equations
$$\Omega_{*}\frac{\partial \Theta}{\partial\tau} + K_{1} \sin{Y} + K_{2} \sin{2 \Theta} \sin{2 Y} = 0\ \ \ \ \ \ \ \ \ \ \ \ \ \ \ \ \ \ \ \ \ \ \ \ \ \ \ \ \ \ \ \eqno(21)$$
$$\Omega_{*}\frac{\partial Y}{\partial\tau}\sin2\Theta + 2 K_{1} \cos{Y} \cos{2 \Theta} - 4 K_{2} \sin{2 \Theta} \cos{2 \Theta} (8 - \cos^{2}{Y}) = 0$$
$$\Theta = \Theta(\tau(\Omega_{*} \tau_{2})),\ \Theta(0) = 0;\ Y = Y(\tau(\Omega_{*} \tau_{2})),\ Y(0) = -\frac{\pi}{2},$$
$$K_{1} = \Omega^{2} \left(\frac{\pi^{2}}{4} - \frac{3 \beta_{1} X}{32}\right);\ K_{2} = \frac{3 \beta_{1} X}{64} \Omega^{2}.$$ 

Using the equation for LPT
$$\cos{Y}=\frac{K_{2}}{K_{1}} \sin{2 \Theta}\left(8 - \cos^{2}{Y}\right),$$
one can get the leading-order approximation of the solution of system $(21)$ in the case of beating between "effective particles"\ far from localization:
$$\Theta_{0}(\tau)=A_{0}\tau,\ \ \Delta_{0}+\Delta_{1}=-\frac{\pi}{2} + \arctan{\left(\frac{2 \sqrt{14}}{7}\tan{\left(\frac{\sqrt{14} K_{1}}{A_{0} \Omega_{*0}}\sin{2 A_0 \tau}\right)}\right)},\eqno(22)$$
$$A_{0}=\frac{\pi}{2}, A_{0} \Omega_{*0} = K_{1}$$
and in case of strongly localized solution:
$$\Theta = A_{*} \sin{\pi \tau}, Y = \frac{\pi}{2}\left(\tau - \frac{1}{2}\right), A_{*} = \arcsin{\frac{K_1}{7 K_2}}, A_{*} \Omega_{*0} = K_{1}.$$

Fig.8  demonstrates that the numerically constructed solution of $(18)$ is in a good agreement with the
yielded approximations.

\begin{center}
a)\includegraphics[width=60mm, height=30mm]{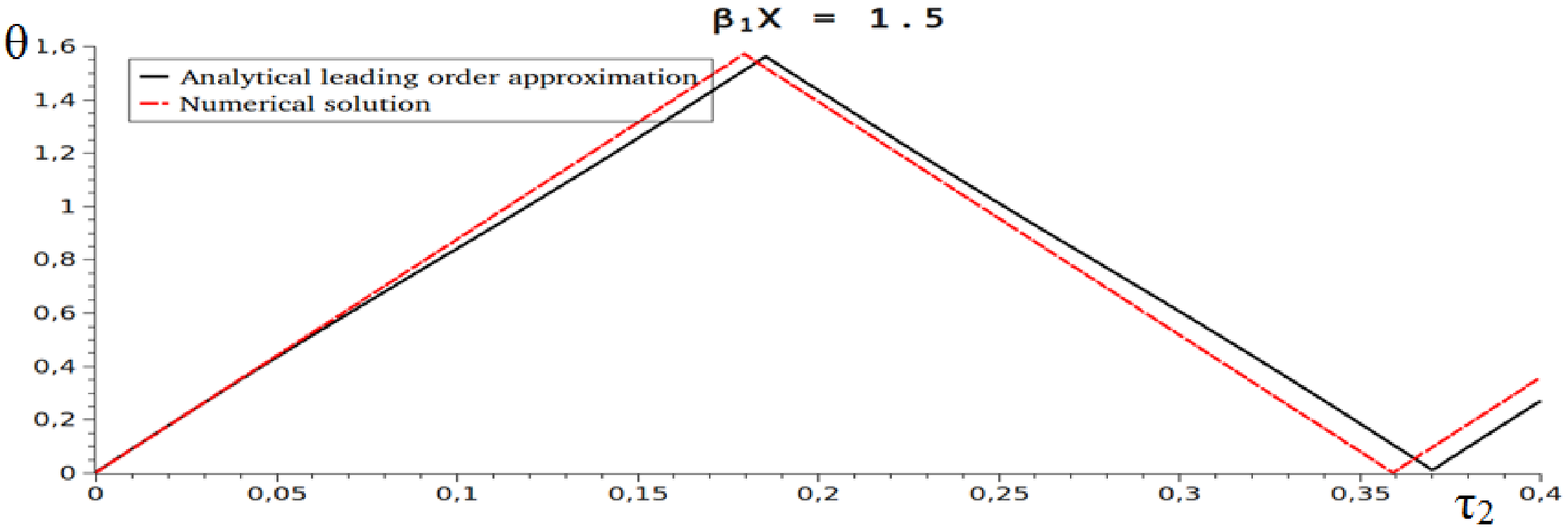} b)\includegraphics[width=60mm, height=30mm]{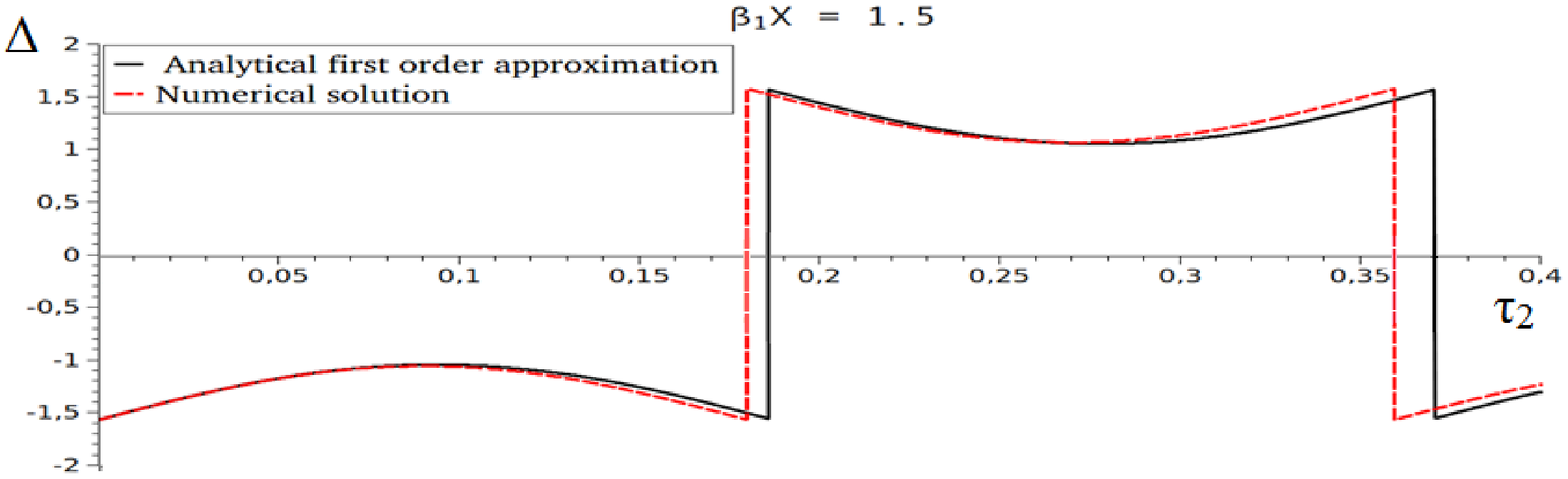}
c)\includegraphics[width=60mm, height=30mm]{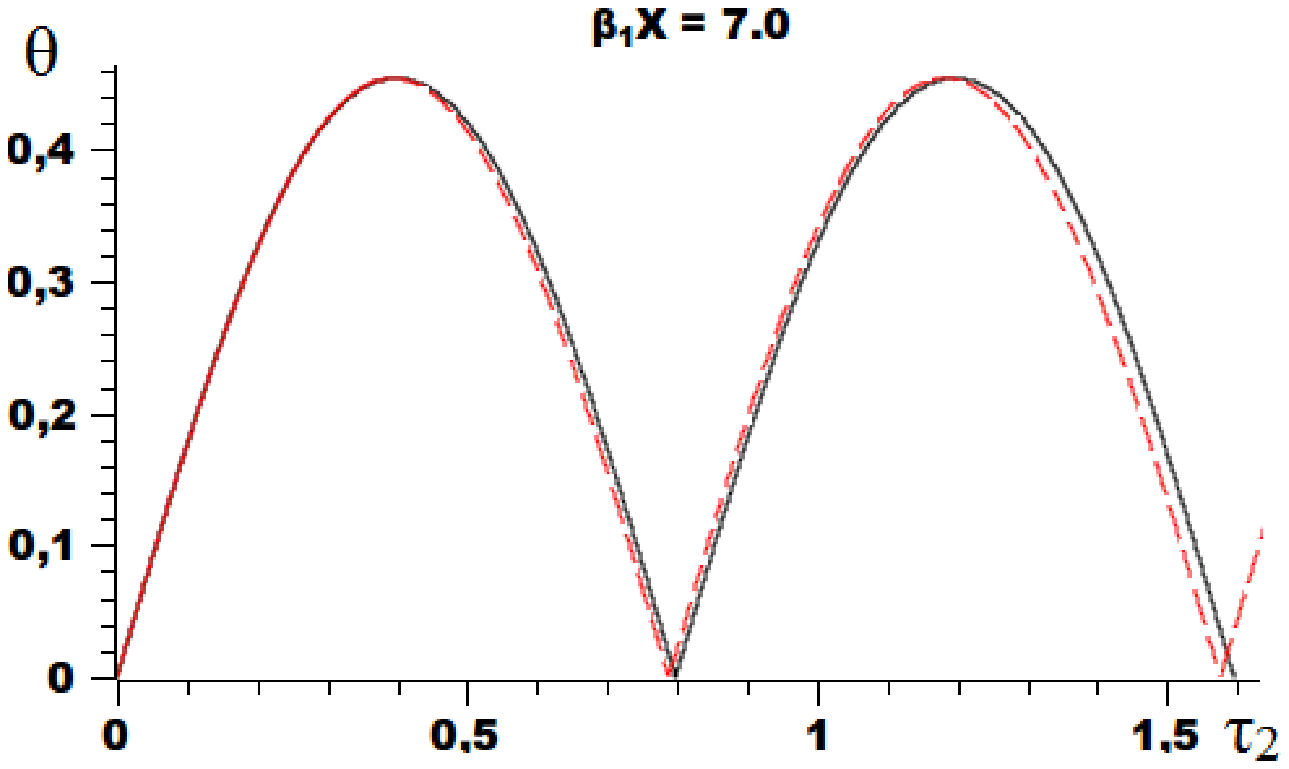} d)\includegraphics[width=60mm, height=30mm]{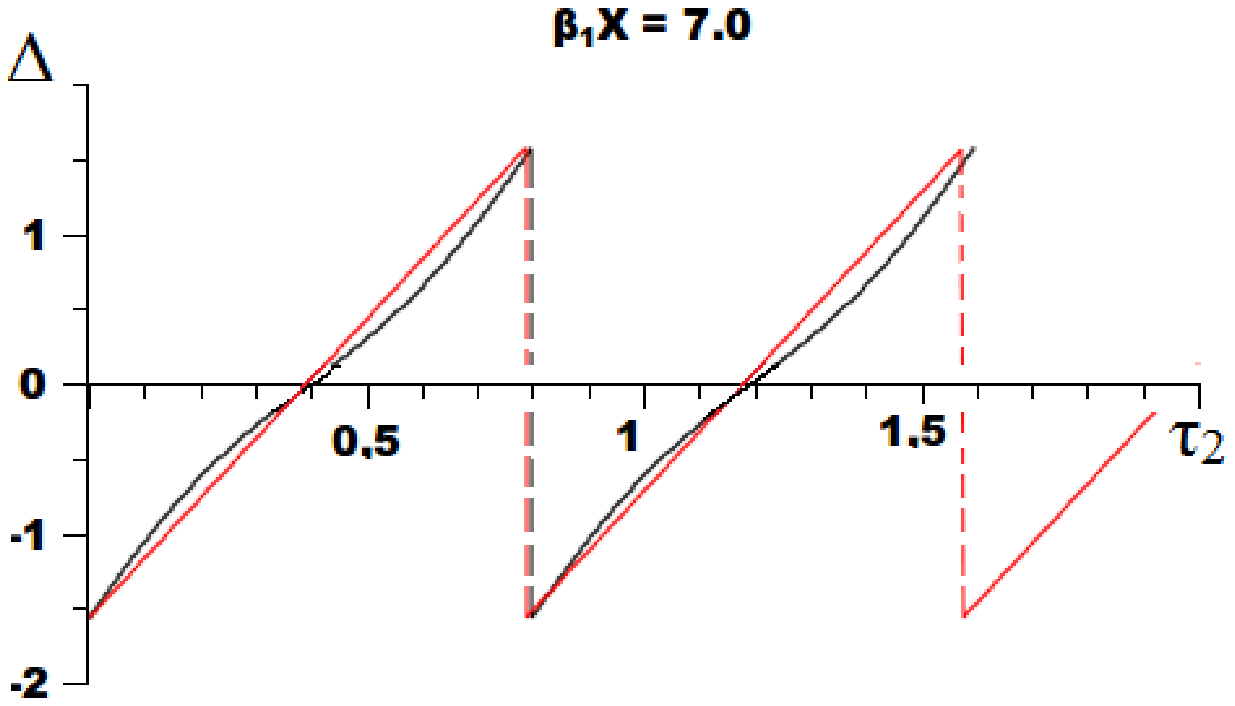}

Fig.8 Comparing of the numerical solution of $(15)$ with the analytical first order approximations in the case of beating between the "effective particles" (a, b) and in the case of strongly localized solution (c, d).
\end{center}

\section{Conclusion}
This study demonstrates that the energy localization phenomenon in the nonlinear oscillatory chains can be described in the framework of the unified concept for the systems with both symmetric and asymmetric interaction potentials.
This concept is based on the notions of "effective particle" and Limiting Phase Trajectory.
It has been shown that asymmetric interaction, typical for molecular systems does not change qualitatively the main features of the energy localization phenomenon observed in the symmetric nonlinear oscillatory chains. 
In fact, the revealed renormalization of the four-order interaction constant substantially simplifies the further analysis.
Besides, this allows us to conclude that asymmetry of the potential leads to reducing the effect of nonlinearity on the dynamical properties of the system.
However, when this renormalized constant becomes small, the higher-orders terms in the expansion of the exact potential energy have to be taken into account. 
The transition to energy localization turns out to be a consequence of the prohibition of the intensive energy exchange between "effective particles". 
The latter is an adequate analogue to the beating phenomenon, well-studied in the systems of two oscillators. 
The analytical results for Limiting Phase Trajectories describing the energy exchange and localization are confirmed with numerical calculations.

 \appendix

 \section{Multiple scale procedure.}
 We employ the multiple scale procedure \cite{Man} in order to consider the processes in the time intervals greatly exceeding $2\pi /\omega_{k}$. Following this procedure, we introduce the time scales:
$$\tau_{0}=t,\ \  \tau_{1}=\varepsilon\tau_{0},\ \  \tau_{2}=\varepsilon^{2}\tau_{0}, \ldots \eqno{(A1)}$$
where the "fast" time $\tau_{0}$  corresponds to the initial time scale of the system, while the slow times $\tau_{1}$, $\tau_{2}$, etc. correspond to the slowly varying envelopes. The envelope functions $\varphi_{k}$ are defined by the relation
$$\Psi_{k}=\varphi_{k}e^{i\omega_{k}\tau_{0}}.$$

$$\frac{d\varphi_{k}}{dt}e^{i\omega_{k} t}-\frac{\alpha\sqrt{2}}{4\sqrt{N}} \sum \limits_{l,m=1}^{N-1} \lambda_{k} \lambda_{l} \lambda_{m} D_{lmk} \left(\varphi_{l}e^{i\omega_{l}t}-\varphi_{l}^{*}e^{-i\omega_{l} t}\right) *$$
$$*\left(\varphi_{m}e^{i\omega_{m}t}-\varphi_{m}^{*}e^{-i\omega_{m} t}\right)+\frac{\beta i}{2N}\sum \limits_{l,m,n=1}^{N-1} \lambda_{k} \lambda_{l} \lambda_{m} \lambda_{n} C_{klmn} \left(\varphi_{l}e^{i\omega_{l}t}-\varphi_{l}^{*}e^{-i\omega_{l} t}\right)*$$
$$*\left(\varphi_{m}e^{i\omega_{m}t}-\varphi_{m}^{*}e^{-i\omega_{m} t}\right) \left(\varphi_{n}e^{i\omega_{n}t}-\varphi_{n}^{*}e^{-i\omega_{n} t}\right)=0.$$

We construct the asymptotic representation of $\varphi_{k}$ in the form
$$\varphi_{k}=\chi_{k,1}+\varepsilon\chi_{k,2}+\varepsilon^{2}\chi_{k,3}+\ldots \eqno{(A2)}$$

The multiple scale expansion based on the relations $(A1)$, $(A2)$ will be used to derive the equations of the leading-order approximation $\chi_{k,1}$.

Terms with $\varepsilon^{0}:$
$$\frac{\partial\chi_{k,1}}{\partial\tau_{0}}=0$$

Terms with $\varepsilon^{1}:$
$$e^{i\omega_{k}\tau_{0}}\left(\frac{\partial\chi_{k,2}}{\partial\tau_{0}}+\frac{\partial\chi_{k,1}}{\partial\tau_{1}}\right)-\frac{\alpha\sqrt{2}}{8}\sum\limits_{l,m=1}^{N-1}\lambda_{k}\lambda_{l}\lambda_{m}D_{lmk} \left(\chi_{l,1}e^{i\omega_{l}\tau_{0}}-\chi_{l,1}^{*} e^{-i\omega_{l}\tau_{0}}\right)*$$
$$* \left(\chi_{m,1}e^{i\omega_{m}\tau_{0}}-\chi_{m,1}^{*} e^{-i\omega_{m}\tau_{0}}\right)=0$$
There is one secular terms in this equation:\ $\frac{\partial\chi_{k,1}}{\partial\tau_{1}}=0 \Rightarrow \chi_{k,1}=\chi_{k,1}(\tau_{2}).$ Then we can get the expression for $\chi_{k,2}$ and use it at the further steps of approximation:
$$\chi_{k,2}=\frac{\alpha\sqrt{2}}{8}\sum\limits_{p,q=1}^{N-1} \lambda_{k}\lambda_{p}\lambda_{q}D_{pqk}\left(\chi_{p,1}\chi_{q,1}\frac{e^{i(\omega_{p}+\omega_{q}-\omega_{k})\tau_{0}}}{i(\omega_{p}+\omega_{q}-\omega_{k})}-\right. $$
$$\left.-2 \chi_{p,1}\chi_{q,1}^{*}\frac{e^{i(\omega_{p}-\omega_{q}-\omega_{k})\tau_{0}}}{i(\omega_{p}-\omega_{q}-\omega_{k})}-\chi_{p,1}^{*}\chi_{q,1}^{*}\frac{e^{-i(\omega_{p}+\omega_{q}+\omega_{k})\tau_{0}}}{i(\omega_{p}+\omega_{q}+\omega_{k})}\right).\eqno{(A3)}$$

Terms with $\varepsilon^{2}:$
$$e^{i\omega_{k}\tau_{0}}\left(\frac{\partial\chi_{k,3}}{\partial\tau_{0}}+\frac{\partial\chi_{k,2}}{\partial\tau_{1}}+\frac{\partial\chi_{k,1}}{\partial\tau_{2}}\right)-$$
$$-\frac{\alpha\sqrt{2}}{8}\sum\limits_{r,s=1}^{N-1}\lambda_{k}\lambda_{r}\lambda_{s}D_{rsk} \left((\chi_{r,2}\chi_{s,1}+\chi_{r,1}\chi_{s,2})e^{i(\omega_{r}+\omega_{s})\tau_{0}}-\right. $$
$$\left.-2(\chi_{r,1}^{*}\chi_{s,2}+\chi_{r,2}^{*}\chi_{s,1})e^{i(\omega_{r}-\omega_{s})\tau_{0}}+(\chi_{r,1}^{*}\chi_{s,2}^{
*}+\chi_{r,2}^{*}\chi_{s,1}^{*})e^{-i(\omega_{r}+\omega_{s})\tau_{0}}\right)+$$
$$+\frac{i \beta}{8}\sum\limits_{l,m,n=1}^{N-1}\lambda_{k}\lambda_{l}\lambda_{m}\lambda_{n} C_{lmnk} \left(\chi_{l,1}e^{i\omega_{l}\tau_{0}}-\chi_{l,1}^{*} e^{-i\omega_{l}\tau_{0}}\right)*$$
$$*\left(\chi_{m,1}e^{i\omega_{m}\tau_{0}}-\chi_{m,1}^{*} e^{-i\omega_{m}\tau_{0}}\right) \left(\chi_{n,1}e^{i\omega_{n}\tau_{0}}-\chi_{n,1}^{*} e^{-i\omega_{n}\tau_{0}}\right)=0$$
One can see that the part of this equation corresponding to the symmetric nonlinear component of the Hamilton function depends only on $\{\chi_{k,1}\}$ and the part corresponding to the asymmetric component can be expressed in $\{\chi_{k,1}\}$ through $(A3)$. Then we integrate this equation with respect to $\tau_{2}$ and obtain equation $(11)$.


\end{document}